\UseRawInputEncoding
\pdfoutput=1

\documentclass[showpacs,nofootinbib,showkeys,eqsecnum,prd,aps,preprint]{revtex4-2}

\usepackage[english]{babel}
\usepackage{comment}

\usepackage{amssymb}
\usepackage{amsmath}
\usepackage{amsfonts}
\usepackage{latexsym}
\usepackage{mathrsfs}
\usepackage{bm}
\usepackage{tensor}
\usepackage{hyperref}
\usepackage{graphicx}

\def\const{\mathop{\rm const}\nolimits\,}
\def\dac{\displaystyle\frac}
\def\dil{\displaystyle\int\limits}

\usepackage{amsthm}

\newtheorem{prop}{Proposition}[section]

\usepackage{diagbox}

\def\const{\mathop{\rm const}\nolimits\,}

\def\dac{\displaystyle\frac}

\def\dil{\displaystyle\int\limits}
\def\{{\lbrace}
\def\}{\rbrace}

\def\Or{{\rm O}}

\begin{document}

\title{Quasiparticles for the one-dimensional nonlocal Fisher-Kolmogorov-Petrovskii-Piskunov equation}

\author{Anton E. Kulagin}
\email{aek8@tpu.ru}
\affiliation{Tomsk Polytechnic University, 30 Lenina av., 634050 Tomsk, Russia}
\affiliation{V.E. Zuev Institute of Atmospheric Optics, SB RAS, 1 Academician Zuev Sq., 634055 Tomsk, Russia}

\author{Alexander V. Shapovalov}
\email{shpv@mail.tsu.ru}
\affiliation{Department of Theoretical Physics, Tomsk State University, Novosobornaya Sq. 1, 634050 Tomsk, Russia}
\affiliation{Laboratory for Theoretical Cosmology, International Centre of Gravity and Cosmos, Tomsk State University of Control Systems and Radioelectronics, 40 Lenina av., 634050 Tomsk, Russia}

\begin{abstract}
We construct quasiparticles-like solutions to the one-dimensional Fisher-Kolmogorov-Petrovskii-Piskunov (FKPP) with a nonlocal nonlinearity using the method of semiclassically concentrated states in the weak diffusion approximation. Such solutions are of use for predicting the dynamics of population patterns. The interaction of quasiparticles stems from nonlocal competitive losses in the FKPP model. We developed the formalism of our approach relying on ideas of the Maslov method. The construction of the asymptotic expansion of a solution to the original nonlinear evolution equation is based on solutions to an auxiliary dynamical system of ODEs. The asymptotic solutions for various specific cases corresponding to various spatial profiles of the reproduction rate and nonlocal competitive losses are studied within the framework of the approach proposed. \\
\end{abstract}



\maketitle

\section{Introduction}
\label{sec:int}

Nonlinear reaction-diffusion equations are used in a number of mathematical models in condensed matter physics, population dynamics, chemical kinetics, etc. Some of phenomena arising in these applications and with long-range interactions are described by the nonlocal nonlinearities within the framework of integro-differential equations. A wide class of such models is related to the Fisher-Kolmogorov-Petrovskii-Piskunov (FKPP) equation with a quadratic nonlocal nonlinearity. The FKPP population equation \cite{fisher37,KPP1937} is the model of a one-species population dynamics.

The solutions to the FKPP equation describe a variety of effects such as clustering \cite{hernandez2004}, spatially localized lumps \cite{paulau2014}, formation of population patterns \cite{murray2001}, propagation of traveling and spiral waves \cite{murray2001,okubo2002}, and steady-state stability \cite{achleitner2013}. The FKPP model is used in studies of cell population dynamics \cite{onofrio2014,fuentes2003,shapob18} including its fractional versions \cite{chu2020,deppman23}. Competitive interactions of microbial populations due to the diffusion of nutrients, the release of toxic substances, chemotaxis, and molecular communications between individuals lead to nonlocal effects that can be effectively described by the generalized nonlocal FKPP equation \cite{fuentes2003,fuentes2004,kenkre2004,lee2001,takeuchi2007}.

The mathematical complexity of the FKPP equation limits the applicable methods for it. The symmetries of this equation were considered in \cite{bluman2013,olver1993}. The symmetry analysis points to the specific cases of the FKPP equation that admit some symmetry groups allowing one to obtain exact solutions. Some travelling wave solutions were found in \cite{volpert09,maruvka07,maruvka06,palencia22,mei2011}. For more complex cases, the studies rely on approximate methods including numerical ones. In \cite{belmonte14}, the collective coordinate method was applied to the FKPP equation (see \cite{sanchez98} and references therein). In \cite{he2006,shakeel13}, the homotopy analysis was used for the traveling wave study. Oscillatory approximate solutions were also obtained in \cite{zhang2023}. The problem under consideration becomes even more complex when the nonlocality comes into play. A necessity for taking into account the nonlocality of the interactions stimulates works where such generalizations are studied (see, e.g., \cite{piva21,hamel13} and references therein).

The attractive feature of the FKPP model is that it predicts the formation and dynamics of population patterns. In the presence of the competition, the population can form the "colonies"\, that are localized around few centres and constitute a pattern. Such "colonies"\, can live long behaving as quasiparticles during their lifetime. The term "quasiparticle"\, originates from the semiclassical study of quantum mechanics. Therefore, it is natural that the formalism of semiclassical approximation can be applied to this problem. The suitable approach that allows one to construct the localized solutions taking into consideration the nonlocal effects was proposed in \cite{fkppshap18}. It is based on the Maslov method \cite{Maslov2,BeD2} that is an effective tool for solving nonlinear problems in quantum mechanics \cite{bagrov1,shapovalov:BTS1,sym2020,belov2007} and related fields \cite{rezaev2007,LST16,shapkul21,shapkul22}.
The objective of this work is to construct the solutions corresponding to few quasiparticles that move over their own trajectories and interact due to ceratin rules dictated by the master equation namely the nonlocal FKPP equation. Actually, these rules are the analogy of the classical equation in the Maslov method in a manner. We limit our mathematical calculations to two quasiparticles and one-dimensional FKPP equation for the clarity of formulae and reasoning since it is sufficient to show the idea of the method and how it describes the interaction of quasiparticles. Nevertheless, the derived formulae can be naturally extended to a more general case of the $n$-dimensional FKPP equation and greater number of quasiparticles.

The paper is organized as follows. In Section \ref{sec:eq}, the FKPP equation under consideration and some notations are given. Then we describe the class of functions where the approximate solutions are to be found. The asymptotics estimates, moments of the desired solution, and asymptotic expansions of the equation coefficients in this class are introduced. In Section \ref{sec:dec}, we decompose the original FKPP equation to two coupled equations each of which is responsible for the population in its own quasiparticle. Also, the equations for moments of this populations are derived in this Section. In Section \ref{sec:red}, the Cauchy problem for the nonlinear nonlocal FKPP equation is reduced to the associated linear problem under certain conditions. The Green functions for the associated linear equations is derived in Section \ref{sec:green}. This Green function generates the approximate solutions to the original nonlinear problem. In Section \ref{sec:sol}, we study the derived asymptotic solutions to the FKPP equation for some particular cases. In Section \eqref{sec:con}, we conclude with some remarks.

\section{The nonlocal 1D FKPP equation}
\label{sec:eq}

Consider the following version of the nonlocal FKPP equation:
\begin{align}
&- u_t(x,t)+D u_{xx}(x,t)+ a(x,t)u(x,t)-\varkappa u(x,t)\int\limits^\infty_{-\infty}b(x,y,t) u(y,t)dy=0.
\label{eq0}
\end{align}
Here and below $u_t=\partial_t u=\partial u/\partial t$, $u_x=\partial_x u=\partial u/\partial x$, $u_{xx}=\partial^2 u/\partial x^2$;
 $a(x,t)$ and  $b(x,y,t)$ are given infinitely
smooth functions increasing, as $|x|\,,|y|\to\infty$, no
faster than the polynomial; $\varkappa \,(>0)$ is a real nonlinearity
parameter.


The term $-\varkappa u(x,t)\int\limits^\infty_{-\infty}b(x,y,t) u(y,t)dy$ stands for the nonlocal competition losses
and is characterized by an influence function $b(x,y,t)$, the coefficient $a(x,t)$ stands for the reproduction rate, and $D$ is the diffusion coefficient.

\subsection{Classes of functions $\mathcal{P}_{s,t}^D$}
\label{sec:class}


Define the common element of the class
$\mathcal{P}_t^D (X(t,D),S(t,D))$ as
\begin{equation}
\label{eq1}
\mathcal{P}_t^D \big(X(t,D),S(t,D) \big)=
  \left\{ \Phi :\Phi(x,t,D)=
  D^{-1/2}\cdot\varphi\left( \frac{\Delta x}{\sqrt D},t,D\right)
\exp \left[ \frac{1}{D}S(t,D)\right]\right\}.
\end{equation}

Here $\Delta x=x-X(t,D)$. The real function $\varphi (\eta,t,D)$
belongs to the Schwarz space $\mathbb{S}$ in the variable
$\eta\in\mathbb{R}^1$, smoothly depends on $t$, and regularly depends
on $\sqrt{D}$ as $D\to 0$. The real functions $S(t,D)$ and $X(t,D)$,
which characterize the class and regularly depend on $\sqrt{D}$ as $D\to 0$, are to be determined. The class $\mathcal{P}_t^D $ is termed as the class of trajectory concentrated function by analogy with \cite{shapovalov:BTS1}.


We will use the contracted notation $\mathcal{P}_{s,t}^D $  instead of $\mathcal{P}_{t}^D (X_s(t,D),S_s(t,D))$ if this does not cause misunderstanding. In order to construct the quasiparticle solutions, we will consider two classes $\mathcal{P}_{s,t}^D$, $s=1,2$. The functions $X_s(t,D)$ correspond to the trajectories of quasiparticles.

\subsubsection{Estimates}

\label{sec:est}

For the functions $u_s$ of the class $\mathcal{P}_{s,t}^D $, $s=1,2$, the following asymptotical estimates hold
\begin{equation}
\label{eq2}
\frac{\|\hat{p}^n \Delta x_s^m u_s\|}{\|u_s\|}= O(D^{(n+m)/2}),
\quad
\frac{\|\hat T(X_s(t,D),t)u_s\|}{\|u_s\|}= O(D),
\end{equation}
where $\Delta x_s=x-X_s(t,D)$, $||\cdot||$ is $L_2$-norm, and
\begin{equation}
\label{T-oper}
\hat{p}=D\partial_x, \quad
\hat T_s(t,D)\equiv \hat T(X_s(t,D),S_s(t,D),t)=D\partial_t+\dot {X}_s(t,D)D\partial_x -\dot {S}_s(t,D).
\end{equation}

In particular, from (\ref{eq2}), it follows that the operators can be estimated as $\hat T_s(t,D)= \widehat{O}(D)$, $\hat {p}= \widehat{O}(\sqrt{D})$ and $\Delta x_s=x-X_s(t,D) = \widehat{O}(\sqrt{D})$. Here, $\widehat O(D^\mu)$ is an operator $\hat F$ such that \[\frac {\|\hat F\phi\|}{\|\phi\|}=O(D^\mu),\quad \phi\in\mathcal{P}_t^D .\]

\subsubsection{Moments}

\label{sec:mom}

To develop a method for constructing the semiclassical solutions to equation (\ref{eq0})
in the class of trajectory concentrated functions (\ref{eq1}), we
assume the existence  of the moments
\begin{eqnarray}
&&\sigma_{u_s}(t,D)=\int\limits^\infty_{-\infty}u_s(x,t,D)dx,\quad
x_{u_s} (t,D)=\frac{1}{\sigma_{u_s}(t,D)}
\int\limits^\infty_{-\infty} xu_s(x,t,D)dx, \cr
&&\alpha_{u_s}^{(l)}(t,D)=
\frac{1}{\sigma_{u_s}(t,D)}\int\limits^\infty_{-\infty}
[x-X_s(t,D)]^l u_s(x,t,D)dx .
\label{momm}
\end{eqnarray}
Here, $s=1,2$, $\sigma_{u_s}$ and $x_{u_s}$  are the zeroth order and first order moments, respectively, and   $\alpha_{u_s}^{(l)}, l\geq 2$ ($l=2,3,\ldots $) is the $l$-th central moment of a 
 function  $u_s(x,t,D)$.

From (\ref{eq2}), we obtain the corresponding estimates for  $\sigma_{u_s}(t)$, $x_{u_s}(t)$, and $\alpha_{u_s}^{(l)}$, $l\geq 2$:
\begin{equation}
\label{estim-alpha}
\sigma_{u_s}(t)= O(1), \quad   x_{u_s}(t) = O(1), \quad
\alpha_{u_s}^{(l)}(t,D)= O(D^{l/2}),
\end{equation}

\subsubsection{Expansions of coefficients}

\label{sec:exp}

Let
\begin{align}
& a(x,t)=\displaystyle\sum_{k=0}^{\infty}\frac{1}{k!}a_{(s)k}(t,D)\Delta x_s^{k},\cr
& b(x,y,t)=\displaystyle\sum_{k,l=0}^{\infty}\frac{1}{k!l!}b_{(s)k,l}(t,D)\Delta x_s^{k}
 \Delta y_s^{l}, \cr
& b(x,y,t)= \displaystyle\sum_{k,l=0}^{\infty}\frac{1}{k!l!}b_{(1,2)k,l}(t,D)\Delta x_1^{k} \Delta y_2^{l},\cr
 &b(x,y,t)=\displaystyle\sum_{k,l=0}^{\infty}\frac{1}{k!l!}b_{(2,1)k,l}(t,D)\Delta x_2^{k}
 \Delta y_1^{l} ,
  \label{expan2}
\end{align}
be formal power series of the functions $a(x,t)$,  $b(x,y,t)$ 
 in equation (\ref{eq0}) in a neighborhood of $X_s(t,D)$. These series will generate asymptotic expansions of the equation operator as $D\to 0$ in view of estimates \eqref{eq2}.
Here, $s, \bar{s}=1,2$, $\Delta y_s=y-X_s(t,D)$,
\begin{align}
& a_{(s)k}(t,D)=\frac {\partial^k a(x,t)}{\partial x^k}\bigg|_{x=X_s(t,D)} , \label{expan1-1}\\
& b_{(s,s)k,l}(t,D)=b_{(s)k,l}(t,D)=\frac {\partial^{k+l} b(x,y,t)}{\partial x^k\partial y^l}
\bigg|_{\begin{subarray}{l}x=X_s(t,D)\\
y=X_s(t,D)\end{subarray}},\cr
& b_{(1,2)k,l}(t,D)=\frac {\partial^{k+l} b(x,y,t)}{\partial x^k\partial y^l}
\bigg|_{\begin{subarray}{l}x=X_1(t,D)\\
y=X_2(t,D)\end{subarray}},\,\,  b_{(2,1)k,l}(t,D)=\frac {\partial^{k+l} b(x,y,t)}{\partial x^k\partial y^l}
\bigg|_{\begin{subarray}{l}x=X_2(t,D)\\
y=X_1(t,D)\end{subarray}}.
\label{expan2-1}
\end{align}
The compact notation is
\begin{align}
& b_{(s,\bar{s})k,l}(t,D)=\frac {\partial^{k+l} b(x,y,t)}{\partial x^k\partial y^l}
\bigg|_{\begin{subarray}{l}x=X_s(t,D)\\
y=X_{\bar{s}}(t,D)\end{subarray}}.
\label{expan2-1a}
\end{align}

Below we use the shorthand notations for convenience: $a_{(s)k}(t,D)=a_{(s) k}(t)$,  $b_{(s)k,l}(t,D)=b_{(s)k,l}(t)$, $b_{(1,2)k,l}(t,D)=b_{(1,2)k,l}(t)$, and
$b_{(2,1)k,l}(t,D)=b_{(2,1)k,l}(t)$.

For the symmetrical influence function, $b(x,y,t)=b(y,x,t)$, we have $b_{(1,2)k,l}(t,D)=b_{(2,1)k,l}(t,D)$.


We impose the technical condition
\begin{equation}
\label{xu}
x_{u_s}(t,D)= X_s(t,D)
\end{equation}
on the functional parameter $X_s(t,D)$ of the functions $u_s(x,t,D)$ belonging to the class $\mathcal{P}_{s,t}^D$. Then from  \eqref{momm} we can write
  \begin{align}
&\alpha_{u_s}^{(0)}(t,D)=1, \quad \alpha_{u_s}^{(1)}(t,D)=0.
\label{momm1}
\end{align}

%
%

\section{Decomposition of the FKPP equation}
\label{sec:dec}

We look for solutions to \eqref{eq0} in the following form:
\begin{align}
u(x,t,D)=u_1(x,t,D)+u_2(x,t,D), \,\, u_s(x,t,D)\in   \mathcal{P}_{s,t}^D,\,\, s=1,2.
\label{dec-1}
\end{align}
Let us the functions $u_s(x,t,D)$ satisfy the equations:
\begin{align}
&-{u}_{s\,t}+D u_{s\,xx}+a(x,t)u_s-\varkappa u_s(x,t,D)\int_{-\infty}^{\infty}
b(x,y,t)\sum_{\bar{s}=1}^{2}u_{\bar{s}}(y,t,D) dy=0,
\label{dec-2}
\end{align}
where $u_{s\,t}=\partial u_s/\partial t$, $u_{s\,xx}=\partial^2 u_s/\partial x^2$.

The summation of the equations \eqref{dec-2} yields the equation \eqref{eq0} for the function $u(x,t,D)$ given by \eqref{dec-1}. The function $u(x,t,D)$ itself does not belong to the class $ \mathcal{P}_{t}^D$ in a general case. Note that the function $u$ \eqref{dec-1} is not a superposition of $u_1$ and $u_2$ since the functions $u_1$ and $u_2$ are interdependent. The function $u_s$ describes the population that belongs to the $s$-th quasiparticle.

The system \eqref{dec-2} is termed as the FKPP decomposition system (FKPPDS).


Next, let us derive the system of equations for the moments \eqref{momm} using \eqref{dec-2}, expansions \eqref{expan2}, \eqref{expan1-1}, \eqref{expan2-1}, and estimates \eqref{estim-alpha}, \eqref{eq2}, \eqref{T-oper}. This system is the generalized Einstein--Ehrenfest system (EES) derived in \cite{fkppshap18} for $u(x,t,D)\in\mathcal{P}_{t}^D$. The EES has the order $M$ if the moments up to the $M$-th order accurate to $O(D^{(n+1)/2})$ are included in it.

\subsubsection{Equations for $\sigma_{u_s}(t,D)$}

For brevity, we will notate $\sigma_s=\sigma_{u_s}(t,D)$, $\alpha^{(2)}_{s}=\alpha^{(2)}_{u_s}$. The time derivative is denoted by, e.g., $\dot{\sigma_s}=d\sigma_s/dt$.

Let us derive the equation for $\sigma_1$. We differentiate the integral \eqref{momm} and substitute $\dot{u}_1$ from \eqref{dec-2}. Next, we substitute the expansions \eqref{expan2}, \eqref{expan1-1}, \eqref{expan2-1} and take into account the estimates  \eqref{estim-alpha}, \eqref{eq2}, \eqref{T-oper}. Finally, we will drop the terms that are estimated as $O(D^{3/2})$ in order to obtain the second order EES that will be used for constructing three terms of an asymptotic expansion for $u(x,t,D)$ later.

Thus, we begin from the following equation for $\sigma_1$:
\begin{align}
&\dot{\sigma_1}(t,D)=\int_{-\infty}^{\infty}{u}_{1\,t}(x,t,D) dx= \int_{-\infty}^{\infty}\big[Du_{1\, xx}+a(x,t)u_1(x,t,D)-\cr
&-\varkappa u_1(x,t,D)\int_{-\infty}^{\infty}b(x,y)\big( u_1(y,t,D)+u_2(y,t,D)\big)dy\big]dx.
\label{dec-sa}
\end{align}
\begin{align}
&\dot{\sigma_1}(t,D)=\int_{-\infty}^{\infty}a(x,t)u_1(x,t,D)dx-\varkappa \int_{-\infty}^{\infty} dx u_1(x,t,D)\int_{-\infty}^{\infty}b(x,y) u_1(y,t,D)dy-\cr
&-\varkappa \int_{-\infty}^{\infty} dx u_1(x,t,D)\int_{-\infty}^{\infty}b(x,y)u_2(y,t,D)dy.
\label{dec-sb}
\end{align}
By analogy, we derive the equation for $\sigma_2$:
\begin{align}
&\dot{\sigma_2}(t,D)=\int_{-\infty}^{\infty}a(x,t)u_2(x,t,D)dx-\varkappa \int_{-\infty}^{\infty} dx u_2(x,t,D)\int_{-\infty}^{\infty}b(x,y) u_1(y,t,D)dy-\cr
&-\varkappa \int_{-\infty}^{\infty} dx u_2(x,t,D)\int_{-\infty}^{\infty}b(x,y)u_2(y,t,D)dy.
\label{dec-sc}
\end{align}
These equations can be written in the unified form:
 \begin{align}
&\dot{\sigma_s}(t,D)=\int_{-\infty}^{\infty}a(x,t)u_s(x,t,D)dx-\varkappa \int_{-\infty}^{\infty} dx u_s(x,t,D)\int_{-\infty}^{\infty}b(x,y)\sum_{\bar{s}=1}^{2} u_{\bar{s}}(y,t,D)dy.
\label{dec-sd}
\end{align}


From \eqref{dec-2}, \eqref{dec-sd}, \eqref{expan1-1},  \eqref{expan2-1a}, \eqref{momm1}, we have
\begin{align}
&\dot{\sigma_s}(t,D)=\sigma_s(t,D)\sum_{k=0}^{\infty}\frac{\alpha_s^{(k)}}{k!}
\Big[a_{(s)k}(t,D)-\varkappa\sum_{l=0}^{\infty}\frac{1}{l!}\sum_{\bar{s}=1}^{2}
b_{(s,\bar{s})k,l}(t,D)\sigma_{\bar{s}}(t,D)\alpha_{\bar{s}}^{(l)}\Big].
\label{mom-sig1a}
\end{align}
Here, $\alpha_{s}^{(k)}=\alpha_{u_s}^{(k)}(t,D)$.

For the prescribed accuracy, we use the expansion \eqref{expan1-1}, \eqref{expan2-1} up to the second power of $\Delta x_s$ inclusive and the moments $\alpha_{s}^{(k)}$ with $k$ of up to 2. The coefficients of the expansion we will write in the following brief form:
\begin{align}
& a_{(s)}(t)\equiv a_{(s)0}(t)=a(X_s(t),t), \,\,
b_{(s)}(t)\equiv b_{(s,s)0,0}(t)= b(X_s(t),X_s(t),t),\cr
& b_{(1,2)}(t)\equiv b_{(1,2)0,0}(t)= b(X_1(t),X_2(t),t),\cr
&b_{(2,1)}(t)\equiv b_{(2,1)0,0}(t)= b(X_2(t),X_1(t),t).
\label{dec-saa}
\end{align}
If $b(x,y,t)=b(x-y,t)=b(y-x,t)$, then $b_{(s)}(t)= b(X_s(t)-X_s(t),t)=b(0,t)$,
$ b_{(1,2)}(t)= b_{(2,1)}(t)=b(X_1(t)-X_2(t),t)$.

The functions $b_{(1,2)}(t), b_{(2,1)}(t)$ and $b_{(1,2)2,0}(t), b_{(2,1)2,0}(t)$, $b_{(1,2)0,2}(t), b_{(2,1)2,0}(t)$ characterize the interaction of quasiparticles moving along the trajectories $x=X_s(t)$.

The resulting equations within the framework of the second order EES read

\begin{align}
& \dot{\sigma}_1=a_{(1)}\sigma_1 +\frac{1}{2} a_{(1)2}\sigma_1\alpha^{(2)}_1-\varkappa\sigma^2_1\Big[b_{(1)} +\frac{1}{2}(b_{(1)0,2} +b_{(1)2,0})\alpha^{(2)}_1 \Big]-\cr
&-\varkappa\sigma_1\sigma_2\Big[b_{(1,2)}+\frac{1}{2}(b_{(1,2)2,0}\alpha^{(2)}_1
+b_{(1,2)0,2}\alpha^{(2)}_2)\Big],
\label{dec-sab}
\end{align}
\begin{align}
& \dot{\sigma}_2=a_{(2)}\sigma_2 +\frac{1}{2} a_{(2)2}\sigma_2\alpha^{(2)}_2-\varkappa\sigma^2_2\Big[b_{(2)} +\frac{1}{2}(b_{(2)0,2} +b_{(2)2,0})\alpha^{(2)}_2 \Big]-\cr
&-\varkappa\sigma_1\sigma_2\Big[b_{(2,1)}+\frac{1}{2}(b_{(2,1)2,0}\alpha^{(2)}_2
+b_{(2,1)0,2}\alpha^{(2)}_1)\Big].
\label{dec-sac}
\end{align}

%

\subsubsection{Equations for $x_{u_s}(t,D)$}
%
Using the relations
\begin{align}
& x_{u_s}(t,D)=x_s(t)=\frac{1}{\sigma_s}\int_{-\infty}^{\infty} xu_s(x,t,D)dx,\,\, \sigma_s=\sigma_s(t,D),\cr
& x=\Delta x_s+X_s(t,D), \,\, \Delta x_s=x-X_s(t,D),
\label{mom-x1a}
\end{align}
we obtain
\begin{align}
& \dot{x}_s(t,D)=-\frac{\dot{\sigma}_u}{\sigma_u}\big(x_s(t,D)-X_s(t,D)\big)+
\frac{1}{\sigma_s}\int_{-\infty}^{\infty}\Delta x_s\dot{u}_s(x,t,D)dx.
\label{mom-x1b}
\end{align}


Subject to the condition \eqref{xu}, we have
\begin{align}
& \dot{x}_s(t,D)=\frac{1}{\sigma_s}\int_{-\infty}^{\infty}\Delta x_s\dot{u}_s(x,t,D)dx.
\label{mom-x1bb}
\end{align}
From \eqref{dec-2}, one readily gets
\begin{align}
& \dot{x}_s=\frac{1}{\sigma_s}\int_{-\infty}^{\infty}\Delta x_s u_s(x,t)dx \Big[ a(x,t)-
 \varkappa\int_{-\infty}^{\infty}b(x,y,t)\sum_{\bar{s}=1}^{2}u_{\bar{s}}(y,t)dy \Big].
\label{mom-x2a}
\end{align}

In view of \eqref{expan1-1},  \eqref{expan2-1a}, \eqref{momm1}, we have
\begin{align}
& \dot{x}_s=\sum_{k=0}^{\infty}\frac{1}{k!}\alpha^{(k+1)}_{s}\Big[
a_{(s)k}(t,D)-\varkappa\sum_{l=0}^{\infty}\frac{1}{l!}\sum_{\bar{s}=1}^{2}b_{(s,\bar{s})k,l}(t,D)
\sigma_{\bar{s}}(t,D)\alpha_{\bar{s}}^{(l)}\Big].
\label{mom-x2b}
\end{align}

For the second order EES, the equations \eqref{mom-x2b} are reduced to
\begin{align}
& \dot{x}_s=\alpha^{(2)}_s\big[a_{(s)1}(t,D)-\varkappa\sum_{\bar{s}=1}^{2}\sigma_{\bar{s}}(t,D)
b_{(s,\bar{s})1,0} (t,D)\big].
\label{mom-x2aa}
\end{align}

%

\subsubsection{Operator $\hat{T}_s$ }

The equation \eqref{mom-x2b} shows that $\dot{x}_s=O(D)$. Then, the term $\dot{X}_s(t,D)\hat{p}$ in \eqref{T-oper} subject to the condition \eqref{xu}
has the estimate $O(D^{3/2})$ that exceeds accuracy of the estimate $\hat{T}_s=O(D)$. Hence, in the subsequent expansions, we will use the following operator instead of $\hat{T}_s$:
\begin{align}
\hat{\mathsf{T}}_s=D\partial_t-\dot{S}_s(t,D),\quad \hat{\mathsf{T}}_s = \widehat{O}(D).
\label{T-oper2}
\end{align}

\subsubsection{Equations for $\alpha^{(n)}_{u_s}(t,D)$, $n\geq 2$.}

\begin{align}
\alpha_s^{(n)}=\frac{1}{\sigma_s}\int_{-\infty}^{\infty}(x-X_s(t,D))^{n}u_s(x,t,D)dx.
\label{mom-alp1}
\end{align}
Using the FKPPDS \eqref{dec-2}, expansions \eqref{expan1-1},
 \eqref{expan2-1a}, conditions  \eqref{xu}, \eqref{momm1}, and equations \eqref{mom-sig1a},
  \eqref{mom-x2b}, we obtain
 \begin{align}
&\dot{\alpha}_s^{(n)}=D n(n-1)  \alpha_s^{(n-2)}+\sum_{k=0}^{\infty}\frac{1}{k!}\big(\alpha_s^{(n+k)}  -n\alpha_s^{(n-1)}\alpha_s^{(k+1)}-\alpha_s^{(n)}\alpha_s^{(k)}\big)\times\cr
&\times\Big[a_{(s)k}(t,D)-\varkappa\sum_{l}^{\infty}\frac{1}{l!}
\sum_{\bar{s}=1}^{2}b_{(s,\bar{s})k,l}(t,D)\alpha_{\bar{s}}^{(l)}\Big],
\label{mom-alp2b}
\end{align}
%
%
For $M=2$, we drop the moments of the order higher than 2 and obtain
\begin{align}
&\dot{\alpha}^{(2)}_s=2 D.
\label{mom-al2}
\end{align}
Then, we have
\begin{align}
&\alpha^{(2)}_s(t,D)=2 D t+ \alpha^{(2)}_s(0).
\label{mom-al22}
\end{align}

The system \eqref{mom-sig1a}, \eqref{mom-x2b}, \eqref{mom-alp2b} for moments of functions $u_1(x,t,D)$ and $u_2(x,t,D)$, which constitute the solution \eqref{dec-1} to the  nonlocal FKPP equation \eqref{eq0}, is the second order EES.

Note that this EES gives the information about the localization of the functions $u_1(x,t,D)$, $u_2(x,t,D)$ and it can be treated as dynamical system of quasiparticles centered over the coordinates $x_{1}(t)$, $x_{2}(t)$. Thus, this dynamical system is the analog of "classical mechanics"\, for the quasiparticles. Since the EES is non-Hamiltonian, the associated "classical"\, system is nonconservative.
%
%
%
%
%

\section{Semiclassical reduction of the FKPPDS to an auxiliary associated linear system}
\label{sec:red}

%
%
%
%
%
%
%
%

Let us introduce the following definition of an asymptotic estimate for the function with respect to its norm:

\begin{align}
&f(x,t,D)=\bar{O}(D^r),\quad r>0 \quad  \Longleftrightarrow \quad \sup_{t\in[0,T]}\parallel f(x,t,D)\parallel=O(D^{r-1/2}),\quad D\to 0.
\label{estbar-1}
\end{align}
The shift of the estimate by $1/2$ is added for convenience since the functions $f\in\mathcal{P}_{s,t}^D$ satisfy $f=\bar{O}(D^0)$ in such notations. Note that we could use the $L_1$-norm instead of the $L_2$. In this case, the proposed estimates would remain valid if the term $1/2$ was missing in \eqref{estbar-1}.

The expansions of $b(x,y,t)$ in \eqref{dec-2} in power series of $\Delta y=y-X(t,D)$ yields

%
%
%
%

\begin{align}
&-{u}_{s\,t}(x,t,D)+ D u_{s\,xx}(x,t,D)+a(x,t)u_{s}(x,t,D)-\cr
&-\varkappa u_s(x,t,D)
\int_{-\infty}^{\infty}\sum_{l=0}^{M}\sum_{\bar{s}=1}^{2}
\frac{1}{l!}b_{[\bar{s}]l}(x,X_{\bar{s}}(t,D),t)\Delta y^l u_{\bar{s}}(y,t,D) dy=\bar{O}(D^{(M+1)/2}),
\label{als-1}
\end{align}
where
\begin{align}
&b_{[\bar{s}]l}(x,X_{\bar{s}}(t,D),t)=\frac{\partial^l b(x,y,t) }{\partial y^l}\Big|_{y=X_{\bar{s}}(t,D)},\cr
&b_{[\bar{s}]}(x,X_{\bar{s}}(t,D),t)=b(x,y,t)\Big|_{y=X_{\bar{s}}(t,D)}.
\label{als-1a}
\end{align}
Having regard to definitions of the moments \eqref{momm} and the notation \eqref{momm1}, we can write the system \eqref{als-1}  as
\begin{align}
&-{u}_{s\,t}(x,t,D)+ D u_{s\,xx}(x,t,D)+a(x,t)u_{s}(x,t,D)-\cr
&-\varkappa u_s(x,t,D)\sum_{l=0}^{M}\sum_{\bar{s}=1}^{2}\sigma_{\bar{s}}(t,D) \frac{1}{l!}b_{[\bar{s}]l}(x,X_{\bar{s}}(t,D),t)\alpha^{(l)}_{\bar{s}}(t,D)=\bar{O}(D^{(M+1)/2}),
\label{als-2a}
\end{align}
or
\begin{align}
&-{u}_{s\,t}(x,t,D)+ D u_{s\,xx}(x,t,D)+a(x,t)u_{s}(x,t,D)-\cr
&-\varkappa u_s(x,t,D)\sum_{\bar{s}=1}^{2}\sigma_{\bar{s}}(t,D)\Big[b_{[\bar{s}]}(x,t,D)+
\sum_{l=2}^{M} \frac{1}{l!}b_{[\bar{s}]l}(x,X_{\bar{s}}(t,D),t)\alpha^{(l)}_{\bar{s}}(t,D)\Big]=\cr
&=\bar{O}(D^{(M+1)/2}).
\label{als-2b}
\end{align}
Note that here $\sigma_{s}=\sigma_{s}^{(M)}$ and $\alpha_s^{(l)}=\alpha_s^{(M)(l)}$ are moments of order $M$, since they are the solution of the Einstein--Ehrenfest system (EES) (of order $M$) which includes moments up to order $M$, inclusive.

We  omit the explicit notation of the order $M$ of the EES in the notation of the moments where it will not lead to confusion.

\begin{align}
\sigma_{s}(t,D,\mathbf{C}),\,\, x_s(t,D,\mathbf{C}),\,\, \alpha_s^{(n)}(t,D,\mathbf{C}),
\,\, n=2,3,\ldots
\label{gsol-1a}
\end{align}

Substituting the general solution  of the EES   of order M
instead of the corresponding moments $\sigma_s=\sigma_{u_s}$, $x_{u_s}$, $\alpha_{s}^{(l)}=\alpha_{u_s}^{(l)}$, $l=2,\ldots ,M$ in \eqref{als-2a} or
\eqref{als-2b},  we obtain a {\it linear system} of partial differential equations
\begin{align}
&\hat{L}_s(x,t,\mathbf{C})v_s(x,t,\mathbf{C})=\cr
&=\big[-\partial_t +D\partial_{xx}+A^{(M)}_s(x,t,\mathbf{C})\big]v_s(x,t,\mathbf{C})=O(D^{(M+1)/2}).
\label{als-3a}
\end{align}
Here
\begin{align}
&A^{(M)}_s(x,t,\mathbf{C})=a(x,t)-\varkappa\sum_{l=0}^{M}\frac{1}{l!}\sum_{\bar{s}=1}^{2}
\sigma_{\bar{s}}^{(M)}(t,\mathbf{C})b_{[\bar{s}]l}(x,x^{(M)}_{\bar{s}}(t,\mathbf{C}),t)
\alpha_{\bar{s}}^{(M)(l)}(t,\mathbf{C}),\cr
& \alpha_s^{(M)(0)}(t,\mathbf{C})=1,\quad \alpha_s^{(M)(1)}(t,\mathbf{C})=0,
\label{als-3b}
\end{align}
and $b_{[\bar{s}]l}(x,x^{(M)}_{\bar{s}}(t,\mathbf{C}),t)$  is clear from \eqref{als-1a}.

The system \eqref{als-3a} can be considered as a family of linear systems of PDEs parametrized by the arbitrary integration constants $\mathbf{C}$.

We shall term the system  \eqref{als-3a} with the coefficients \eqref{als-3b} the associated linear system for the decomposed FKPP system \eqref{dec-2} of order $M$.

Из \eqref{mom-sig1a} и \eqref{als-3b} запишем
%
\begin{align}
&A_s^{(M)}(x,t,\mathbf{C})-\frac{\dot{\sigma}_s^{(M)}(t,\mathbf{C})}{\sigma_s^{(M)}(t,\mathbf{C})}=
\sum_{k=0}^{M}\frac{1}{k!}\Big(\Delta x_s^k-\alpha_s^{(M)(k)} \Big)\Big[
a_{(s)k}-\cr
&-\varkappa\sum_{l=0}^{M-k}\frac{1}{l!}\sum_{\bar{s}=1}^{2}
b_{(s,\bar{s})k,l}\sigma_{\bar{s}}^{(M)}\alpha_{\bar{s}}^{(M)(l)}\Big].
\label{als-3c}
\end{align}
Here, $a_{(s)k}=a_{(s)k}(t,\mathbf{C})$ follows from \eqref{expan1-1} where
we put $x=x_s(t,\mathbf{C})$, $b_{(s,\bar{s})k,l}=b_{(s,\bar{s})k,l}(t,\mathbf{C})$
follows from \eqref{expan2-1a} where $x=x_s(t,\mathbf{C})$, $y=x_{\bar{s}}(t,\mathbf{C})$,
and  $\sigma_s^{(M)}=\sigma_s^{(M)}(t,\mathbf{C})$, $x_s=x_s(t,\mathbf{C})$, and
 $\alpha^{(M)(k)}_{s}=\alpha^{(M)(k)}_{s}(t,\mathbf{C})$ is the general solution to the
 EE system \eqref{mom-sig1a}, \eqref{mom-x2b}, and \eqref{mom-alp2b} of the order $M$.

We rewrite \eqref{als-3c} as
\begin{align}
&A_s^{(M)}(x,t,\mathbf{C})=\frac{\dot{\sigma}_s^{(M)}(t,\mathbf{C})}{\sigma_s^{(M)}(t,\mathbf{C})}
+\big[a_{(s)1}-\varkappa\sum_{l=0}^{M}\frac{1}{l!}\sum_{\bar{s}=1}^{2}b_{(s,\bar{s})1,l}\sigma^{(M)}_{\bar{s}}
\alpha^{(M)(l)}_{\bar{s}} \big]\Delta x_s+\cr
&+\sum_{k=2}^{M}\frac{1}{k!}\Big(\Delta x_s^k-\alpha_s^{(M)(k)} \Big)\Big[
a_{(s)k}-\varkappa\sum_{l=0}^{M-k}\frac{1}{l!}\sum_{\bar{s}=1}^{2}
b_{(s,\bar{s})k,l}\sigma_{\bar{s}}^{(M)}\alpha_{\bar{s}}^{(M)(l)}\Big].
\label{als-3c}
\end{align}

Let us write the operator $\hat{L}_s(x,t,\mathbf{C})$ of \eqref{als-3a} as
\begin{align}
&D\hat{L}_s(x,t,\mathbf{C})=-\hat{\mathsf{T}}_s(t,\mathbf{C})-\dot{S}_s(t,D)+
\hat{p}^2+D A^{(M)}(x,t,\mathbf{C}).
\label{als-4a}
\end{align}
Substituting \eqref{als-3c} and \eqref{mom-x2b}, 
we can write \eqref{als-4a} as
\begin{align}
&D\hat{L}_s(x,t,\mathbf{C})=-\hat{\mathsf{T}}_s(t,\mathbf{C})-\dot{S}_s(t,D)+ \hat{p}^2+\cr
&+D\frac{\dot{\sigma}_s^{(M)}(t,\mathbf{C})}{\sigma_s^{(M)}(t,\mathbf{C})}+
\Big[a_{(s)1}-\varkappa\sum_{l=0}^{M-1}\frac{1}{l!}\sum_{\bar{s}=1}^{2}b_{(s,\bar{s})1,l}
\sigma^{(M)}_{\bar{s}}\alpha^{(M)(l)}_{\bar{s}} \Big] D\Delta x_s+\cr
&+D\sum_{k=2}^{M}\frac{1}{k!}\big( \Delta x_s^{k} -\alpha_s^{(M)(k)} \big)
\Big[a_{(s)k}-\varkappa\sum_{l=0}^{M-k}\frac{1}{l!}\sum_{\bar{s}=1}^{2}
b_{(s,\bar{s})k,l}\sigma_{\bar{s}}^{(M)}\alpha_{\bar{s}}^{(M)(l)}\Big],
\label{als-4b}
\end{align}
 where $\hat{\mathsf{T}}_s(t,\mathbf{C})=$  $D\partial_t-\dot{S}_s(t,D)$
   according to  \eqref{T-oper2} and  $\hat{\mathsf{T}}_s(t,\mathbf{C})= \widehat{O}(D)$, and also $\hat{p}=D\partial_x= \widehat{O}(\sqrt{D})$, $\Delta x_s= \widehat{O}(\sqrt{D})$.

For brevity of notation, we omit the explicit dependence on $D$ where this does not cause  confusion.

Denote by $h^{(M)}_{(s)k}=h^{(M)}_{(s)k}(t,\mathbf{C})$ the function
\begin{align}
&h^{(M)}_{(s)k}=\sum_{l=0}^{M-k}\frac{1}{l!}\sum_{\bar{s}=1}^{2}b_{(s,\bar{s})k,l}
\sigma_{\bar{s}}^{(M)} \alpha^{(M)(l)}_{\bar{s}}.
\label{als-4ba}
\end{align}
Then, using \eqref{als-4ba}, we rewrite \eqref{als-4b} in a more compact way
\begin{align}
&D\hat{L}_s(x,t,\mathbf{C})=-\hat{\mathsf{T}}_s(t,\mathbf{C})-\dot{S}_s(t,D)+ \hat{p}^2+\cr
&+D\frac{\dot{\sigma}_s^{(M)}(t,\mathbf{C})}{\sigma_s^{(M)}(t,\mathbf{C})}+
\Big(a_{(s)1}-\varkappa h^{(M)}_{(s)1}\Big) D\Delta x_s+\cr
&+D\sum_{k=2}^{M}\frac{1}{k!}\big( \Delta x_s^{k} -\alpha_s^{(M)(k)} \big)
\Big(a_{(s)k}-\varkappa h^{(M)}_{(s)k}\Big),
\label{als-4bb}
\end{align}

The function $S_s(t,D)$ is a free parameter of the class (\ref{eq1}).
Let us choose $S_s(t,D)$ such that
\begin{equation}
\dot{S}_s(t,D)=D\frac{\dot{\sigma}_s^{(M)}(t,\mathbf{C})}{\sigma_s^{(M)}(t,\mathbf{C})}.
\label{s-func-1}
\end{equation}
Then we have
\begin{equation}
\exp{\Big(\frac{1}{D}\big(S_s(t,D)-S_s(0,D)\big)\Big)}=
\exp{\Big(\int_0^t\frac{\dot{\sigma}_s^{(M)}(\tau,\mathbf{C})}{\sigma_s^{(M)}(\tau,\mathbf{C})}d\tau\Big)}=
\frac{\sigma_s^{(M)}(t,\mathbf{C})}{\sigma_s^{(M)}(0,\mathbf{C})}.
\label{s-func-2}
\end{equation}
From \eqref{mom-x2b}, we see that $\dot{x}_s^{(M)}(t,\mathbf{C})= O(D)$,
and then $\dot{x}_s^{(M)}(t,\mathbf{C})\hat{p}= \widehat{O}(D^{3/2})$.

In view of \eqref{s-func-2}and  estimates of the operators contained in \eqref{als-4b}, we can represent the operator $D\hat{L}_s(x,t,\mathbf{C})$  as a decomposition
\begin{align}
&D\hat{L}_s(x,t,\mathbf{C})=\hat{L}_{s(0)}^{(M)}(t,\mathbf{C})+
\hat{L}_{s(1)}^{(M)}(t,\mathbf{C})+\hat{L}_{s(2)}^{(M)}(t,\mathbf{C})+\ldots ,
\label{als-5a}
\end{align}
where $\hat{L}_{s(0)}^{(M)}(t,\mathbf{C})=\widehat{O}(D)$,
$\hat{L}_{s(1)}^{(M)}(t,\mathbf{C})=\widehat{O}(D^{3/2})$, $\hat{L}_{s(2)}^{(M)}(t,\mathbf{C})=\widehat{O}(D^2)$, $\ldots \,$,
$\hat{L}_{s(k)}^{(M)}(t,\mathbf{C})=\widehat{O}(D^{(k+2)/2}$), $k=1,2,\ldots , M$, and
\begin{align}
&\hat{L}_{s(0)}^{(M)}(t,\mathbf{C})=-\hat{\mathsf{T}}_s(t,\mathbf{C})+\hat{p}^2,
\label{L0}\\
&\hat{L}_{s(1)}^{(M)}(t,\mathbf{C})=\Big(a_{(s)1}-\varkappa h^{(M)}_{(s)1} \Big)D\Delta x_s \label{L1}\\
&\hat{L}_{s(2)}^{(M)}(t,\mathbf{C})=\frac{1}{2}\Big(a_{(s)2}-\varkappa h^{(M)}_{(s)2}\Big)
D\big(\Delta x_s^2-\alpha_s^{(M)(2)}\big), \label{L2}\\
&\qquad \ldots  \cr
&\hat{L}_{s(k)}^{(M)}(t,\mathbf{C})=\frac{1}{k!}\Big(a_{(s)k}-\varkappa h^{(M)}_{(s)k} \Big)
 D\big(\Delta x_s^k-\alpha_s^{(M)(k)}\big).
\label{Lk}
\end{align}

Note that the choice of operators \eqref{L0} -- \eqref{Lk} is not unique.

We shall seek the trajectory concentrated approximate solutions
$v_s(x,t,D,\mathbf{C})\in\mathcal{P}_{s,t}^D\big(x_s^{(M)}(t,D,\mathbf{C}),S(t,D)\big)$
to the linear associated system \eqref{als-3a}, \eqref{als-3b} as the asymptotic series:
\begin{align}
&v_s(x,t,D,\mathbf{C})=v^{(0)}_s(x,t,D,\mathbf{C})+\sqrt{D}v^{(1)}_s(x,t,D,\mathbf{C})+
D v^{(2)}_s(x,t,D,\mathbf{C})+\ldots
\label{als-5b}
\end{align}
Here, $\mathcal{P}_{s,t}^D$ is given  in \eqref{eq1},
$v^{(k)}_s(x,t,D,\mathbf{C})\in $ $\mathcal{P}_{s,t}^D\big(x_s^{(M)}(t,D,\mathbf{C}),S(t,D)\big)$, and
all the functions $v^{(k)}_s(x,t,D,\mathbf{C})$ have the same estimate
$\bar{O}(D^{m})$ for some $m$  as $D\to 0$ in the sense of definition \eqref{estbar-1}.
%

Substituting the formal expansions \eqref{als-5a}, \eqref{als-5b} into the linear equation \eqref{als-3a}, we can obtain
\begin{align}
& D\hat{L}_s v_s(x,t,D,\mathbf{C})=\bar{O}(D^{(M+3)/2}).
\label{als-5c}
\end{align}
 Equating the terms having the same estimate in the parameter $D$
 in the sense of definition \eqref{estbar-1}, we get the following
system of recurrent equations:
\begin{align}
& \hat{L}_{s(0)}^{(M)}(t,\mathbf{C})v^{(0)}_s(x,t,D,\mathbf{C})=0, \label{als-6a}\\
&\sqrt{D} \hat{L}_{s(0)}^{(M)}(t,\mathbf{C})v^{(1)}_s(x,t,D,\mathbf{C})+
\hat{L}_{s(1)}^{(M)}(t,\mathbf{C})v^{(0)}_s(x,t,D,\mathbf{C})=0,   \label{als-6b}\\
& D\hat{L}_{s(0)}^{(M)}(t,\mathbf{C})v^{(2)}_s(x,t,D,\mathbf{C})+
\sqrt{D}\hat{L}_{s(1)}^{(M)}(t,\mathbf{C})v^{(1)}_s(x,t,D,\mathbf{C})+\cr
&+\hat{L}_{s(2)}^{(M)}(t,\mathbf{C})v^{(0)}_s(x,t,D,\mathbf{C})=0,   \label{als-6c}\\
&\qquad \qquad \ldots \cr
&D^{k/2}\hat{L}_{s(0)}^{(M)}(t,\mathbf{C})v^{(k)}_s(x,t,D,\mathbf{C})
+\sum_{j=1}^{k}D^{(k-j)/2}\hat{L}_{s(j)}^{(M)}(t,\mathbf{C})v^{(k-j)}_s(x,t,D,\mathbf{C})=0. \label{als-6d}
\end{align}
These equations in compact notation have the form
\begin{align}
&\sum_{j=0}^{k}D^{(k-j)/2}\hat{L}_{s(j)}^{(M)}(t,\mathbf{C})v^{(k-j)}_s(x,t,D,\mathbf{C})=0,\quad
k=0,1, \ldots , M.
 \label{als-7}
\end{align}

%
\[
v^{(k)}_s(x,t,D,\mathbf{C})
=- \big(\hat{L}_{s(0)}^{(M)}\big)^{-1}(t,\mathbf{C})  \sum_{j=1}^{k}D^{-j/2}\hat{L}_{s(j)}^{(M)}(t,\mathbf{C})v^{(k-j)}_s(x,t,D,\mathbf{C}),\,\, k>0.
\]

\subsection{The Cauchy problem}

Let us consider the Cauchy problem for the FKPPDS \eqref{dec-2} in the class \eqref{eq1}:
\begin{align}
&u_s(x,t,D)\big|_{t=0}=\varphi_s(x,D),\,\, s=1,2,\cr
&u_s(x,t,D)\in \mathcal{P}^D_{s,t}=\mathcal{P}^D_{t}\left(X_s(t,D),S_s(t,D)\right),\quad
\varphi_s(x,D)\in \mathcal{P}^D_{s,0}=\mathcal{P}^D_{s,t}\big|_{t=0}.
\label{cauch-1a}
\end{align}
The initial condition \eqref{cauch-1a} generates the Cauchy problem for the EES
\eqref{mom-sig1a}, \eqref{mom-x2b}, \eqref{mom-alp2b}
\begin{align}
&\sigma_s(t,D)\big|_{t=0}=\sigma_s(D,\varphi_s),\,\,
x_s(t,D)\big|_{t=0}=x_s(D,\varphi_s),\,\,
\alpha_s^{(l)}(t,D)\big|_{t=0}=\alpha_s(D,\varphi_s), \, l=2,3,...
\label{cauch-1c}
\end{align}
Here, $\sigma_s(D,\varphi_s)$, $x_s(D,\varphi_s)$, $\alpha_s(D,\varphi_s)$
are given by \eqref{momm} for $t=0$.

Let the integration constants $\mathbf{C}$ in the general solution \eqref{gsol-1a} be determined by the following algebraic conditions:
\begin{align}
&\sigma_s(t,D,\mathbf{C})\big|_{t=0}=\sigma_s(D,\varphi_s),\,\,
x_s(t,D,\mathbf{C})\big|_{t=0}=x_s(D,\varphi_s),\cr
&\alpha_s^{(l)}(t,D,\mathbf{C})\big|_{t=0}=\alpha_s(D,\varphi_s),\,\, l=2,3,\ldots
\label{alg-1a}
\end{align}
Then, the integration constants $\mathbf{C}$ are the functionals depending on both initial functions $\vec{\varphi}=(\varphi_1,\varphi_2)$:
\begin{align}
&\mathbf{C}=\mathbf{C}[\vec{\varphi}].
\label{alg-1b}
\end{align}

The moments of $u_s$ that are determined by EES and the particular solutions to EES with integration constants given by \eqref{alg-1b} are equal due to the uniqueness of the Cauchy problem solution
\begin{align}
&\sigma_s(t,D)[u_s]=\sigma_s(t,D,\mathbf{C}[\vec{\varphi}]),\,\,
 x_s(t,D)[u_s]= x_s(t,D, \mathbf{C}[\vec{\varphi}]), \cr
 & \alpha_s^{(n)}(t,D)[u_s]=
 \alpha_s^{(l)}(t,D,\mathbf{C}[\vec{\varphi}]) ,\,\, l=2,3,\ldots
\label{cauch-1bb}
\end{align}

Then, the following proposition holds.
\begin{prop}
\label{teo1}
Let $u_s(x,t,D)\in  {\mathcal{P}}_{s,t}^D = {\mathcal{P}}_t^D\left(X_s(t,D),S_s(t,D)\right)$, $s=1,2$, be solutions to the equation \eqref{dec-2} and $v_s^{(k)}(x,t,D,\mathbf{C})$, $s=1,2$, be solutions to the equations \eqref{als-7}.

Let $\mathbf{C}$ be the integration constants determined by
\eqref{alg-1b} and the initial conditions $u_s(x,t,D)\Big|_{t=0}=\varphi_{s}(x,D)$, i.e. $\mathbf{C}=\mathbf{C}[\vec{\varphi}]$, where
\begin{equation}
\varphi_{s}(x,D)=\sum_{k=0}^{M}D^{k/2}v_s^{(k)}(x,t,D,\mathbf{C})\Big|_{t=0}, \quad s=1,2.
\label{demo1}
\end{equation}
\par

Then,
\begin{align}
&v_s(x,t,D,\mathbf{C}[\vec{\varphi}])=
\sum_{k=0}^{M}D^{k/2}v_s^{(k)}(x,t,D,\mathbf{C}[\vec{\varphi}]), \,\, s=1,2,
\label{cauch-2a}
\end{align}
 satisfy the equations \eqref{dec-2} with the right-hand side accuracy of $\bar{O}(D^{(M+1)/2})$
and
\begin{equation}
v_s(x,t,D,\mathbf{C}[\vec{\varphi}])-u_s(x,t,D)=\bar{O}(D^{(M+1)/2}), \quad s=1,2.
\label{cauch-2b}
\end{equation}

  \end{prop}

%
%

\section{The Green function}
\label{sec:green}

The Green function $G_{s(0)}^{(M)}(x,y,t,t_0,D,\mathbf{C})$ is determined by the conditions
\begin{align}
&\hat{L}_{s(0)}^{(M)}G_{s(0)}^{(M)}(x,y,t,t_0,D,\mathbf{C})=0,\quad t>t_0, \cr
&G_{s(0)}^{(M)}(x,y,t,t_0,D,\mathbf{C})=\delta(x-y), \quad x,y \in\mathbb{R}^1.
\label{green-1}
\end{align}
For $k=1,2, \ldots$, we can write the recurrent formula for
$v^{(k)}_s(x,t,D,\mathbf{C})$ according to Duhamel's principle:
\begin{eqnarray*}
&&v^{(k)}_s(x,t,D,\mathbf{C})
=\int_{-\infty}^{\infty} G_{s(0)}^{(M)}(x,y,t,t_0,D,\mathbf{C})v^{(k)}_s(y,t_0,D,\mathbf{C})dy+ \cr &&+\int_{t_0}^{t}d\tau \int_{-\infty}^{\infty} G_{s(0)}^{(M)}(x,y,t,\tau,D,\mathbf{C})
 \sum_{j=1}^{k} D^{-(j+2)/2}\hat{L}_{s(j)}^{(M)}(\tau  ,\mathbf{C})v^{(k-j)}_s(y,\tau,D,\mathbf{C})dy.
\label{geen-2}
\end{eqnarray*}

For shorthand  notation we denote  $G_{s(0)}^{(M)}(x,y,t,t_0,D,\mathbf{C})=G_s(x,y,t,t_0)$ for an initial $t_0$ and write equations \eqref{green-1} as
\begin{align}
&\Big[ -\big(D\partial_t -\dot{S}_s\big)+(D\partial_x)^2\Big]G_{s}(x,y,t,t_0)=0,\quad t>t_0, \cr
&G_{s}(x,y,t_0,t_0)=\delta(x-y).
\label{green-1a}
\end{align}
The direct and inverse Fourier transforms for a function $f(x)$ are
\begin{align}
&F(k)=\frac{1}{\sqrt{2\pi}}\int_{-\infty}^{\infty}e^{-ikx}f(x) dx,\quad
f(x)=\frac{1}{\sqrt{2\pi}}\int_{-\infty}^{\infty}e^{ikx}F(k) dk.
\label{four-1a}
\end{align}
For the Fourier transform $\tilde{G}_{s}(k,y,t,t_0)$ of $G_{s}(x,y,t,t_0)$, from \eqref{four-1a} we find
\begin{align}
&\Big[ -\big(D\partial_t -\dot{S}_s\big)-k^2 D^2\Big]\tilde{G}_{s}(k,y,t,t_0)=0,\cr
&\tilde{G}_{s}(k,y,t_0,t_0)=\frac{1}{\sqrt{2\pi}}e^{-ik y},
\label{green-1b}
\end{align}
whence it follows
\begin{align}
&\tilde{G}_{s}(k,y,t,t_0)=\frac{1}{\sqrt{2\pi}}\exp\Big[-iky -k^2D\Delta t+\frac{\Delta S_s}{D} \Big],
\label{green-1c}
\end{align}
where  $\Delta t=t-t_0$,  $\Delta S_{s}=S_{s}(t)-S_{s}(t_0)$.
The inverse Fourier transform of \eqref{green-1c} reads
\begin{align}
&G_{s}(x,y,t,t_0)=\frac{1}{2\sqrt{\pi\Delta t D}}\exp\Big[-\frac{(x-y)^2}{4D \Delta t}+\frac{\Delta S_{s}}{D} \Big].
\label{green-2}
\end{align}
Note: the Gaussian $\delta$-pulse is
\[
\delta(x)=\lim_{\varepsilon\to 0}\frac{1}{2\sqrt{\pi \varepsilon}}e^{-\frac{x^2}{4\varepsilon}},\quad \varepsilon >0.
\]


%
%
%
%
%

\section{Asymptotic solutions}

\label{sec:sol}

%

In view of \eqref{cauch-2b}, the roughest asymptotic solution to \eqref{eq0} corresponds to $M=0$. It is the leading term of asymptotics with right-hand side accuracy of $\bar{\Or}(D^{3/2})$. In order to construct such solution, we need the solutions to the EES of the zeroth order. Such EES do not describes the dynamics of quasiparticles in the $x$-space since it does not involves the first moment of the solution. Moreover, for $M=1$, the EES system contains the trivial equation $\dot{x}_s=0$ since it is clear from \eqref{mom-x2b} that $\dot{x}_s=\Or(D)$ that exceeds the accuracy $\Or(D^{1/2})$ of the first order EES. Hence, the roughest asymptotic solution that catches the dynamics of quasiparticles is $v_s(x,t)=v_s^{(0)}(x,t)+D^{1/2}v_s^{(1)}(x,t)+Dv_s^{(2)}(x,t)$ corresponding to the EES of the second order given by \eqref{dec-sab}, \eqref{dec-sac}, \eqref{mom-x2aa}, \eqref{mom-al2}. Let us construct this asymptotic solution under simplifying assumption $v_s^{(0)}(x,t)\big|_{t=0}=\varphi_s(x)$. Then, we have
\begin{equation}
v_s^{(1)}(x,t)\big|_{t=0}=v_s^{(2)}(x,t)\big|_{t=0}=0.
\label{nontr1}
\end{equation}
Under this assumption, the formula \eqref{geen-2} yields
\begin{equation}
\begin{gathered}
v^{(0)}_s(x,t)=\int_{-\infty}^{\infty} G_{s(0)}^{(M)}(x,y,t,0) \varphi_s(y)dy,\\
v^{(1)}_s(x,t)=\int_{0}^{t}d\tau \int_{-\infty}^{\infty} G_{s(0)}^{(M)}(x,y,t,\tau) D^{-3/2}\hat{L}_{s(1)}^{(M)}(\tau)v^{(0)}_s(y,\tau)dy,\\
v^{(2)}_s(x,t)=\int_{0}^{t}d\tau \int_{-\infty}^{\infty} G_{s(0)}^{(M)}(x,y,t,\tau) \left(D^{-3/2}\hat{L}_{s(1)}^{(M)}(\tau)v^{(1)}_s(y,\tau)+D^{-4/2}\hat{L}_{s(2)}^{(M)}(\tau)v^{(0)}_s(y,\tau)\right)dy.
\end{gathered}
\label{nontr2}
\end{equation}
Let the quasiparticles have a gaussian initial wave packets:
\begin{equation}
\varphi_s(x)=N_s D^{-1/2}\exp\!\left(-\dac{(x-x_s(0))^2}{2D\gamma_s^2}\right), \qquad \gamma_s>0.
\label{nontr3}
\end{equation}
Then, we have
\begin{equation}
\begin{gathered}
v^{(0)}_s(x,t)=N_s D^{-1/2}\exp\!\left[\dac{\Delta S_s}{D}\right]\dac{\gamma_s}{\sqrt{2t + \gamma_s^2}}\exp\!\left[-\dac{(x-x_s(0))^2}{4D t + 2 D\gamma_s^2}\right],
\end{gathered}
\label{resh0}
\end{equation}

\begin{equation}
\begin{gathered}
v^{(1)}_s(x,t)=\dac{N_s \gamma_s}{D\sqrt{ (\gamma_s^2+2t)^3}}\exp\!\left[\dac{ S_s(t)-S_s(0)}{D}\right]\exp\!\left[-\dac{(x-xs(0))^2}{4Dt+2D\gamma_s^2}\right]   \times \\
\times\dil_0^{t} d\tau  k_{(s)1}(\tau)\Big((\gamma_s^2+2\tau) \big(x-x_s(0)\big)-(\gamma_s^2+2t)\big(x_s(\tau)-x_s(0)\big)\Big),
\end{gathered}
\label{resh1}
\end{equation}

\begin{equation}
\begin{gathered}
v^{(2)}_s(x,t)=N_s\exp\!\left[\dac{ S_s(t)-S_s(0)}{D}\right]\exp\!\left[-\dac{(x-xs(0))^2}{4Dt+2D\gamma_s^2}\right] \dac{\gamma_s}{D^{3/2}(\gamma_s^2+2t)^{5/2}}\times\\
\times \Bigg[\dil_0^{t} d\tau_2 k_{(s)1}(\tau_2) \dil_0^{\tau_2}  d\tau_1 k_{(s)1}(\tau_1)\Bigg\{ (\gamma_s^2+2\tau_2)(\gamma_s^2+2\tau_1)(x-x_s(0))^2-\\
- (\gamma_s^2+2t)\Big((\gamma_s^2+2\tau_2)\big(x_s(\tau_1)-x_s(0)\big)+(\gamma_s^2+2\tau_1)\big(x_s(\tau_2)-x_s(0)\big)\Big)\big(x-x_s(0)\big)+\\
+(\gamma_s^2+2t)\bigg(2D\gamma_s^2 (\tau_1,\tau_2)t+4D t\tau_1-2D\gamma_s^2\tau_2-4D\tau_1\tau_2+\\
+(\gamma_s^2+2t)\big(x_s(\tau_2)-x_s(0)\big)\big(x_s(\tau_1)-x_s(0)\big)\bigg)\Bigg\}+\\
+\dil_0^{t} d\tau_2 k_{(s)2}(\tau_2)\Bigg\{(\gamma_s^2+2\tau_2)^2(x-x_s(0))^2- 2(\gamma_s^2+2t)(\gamma_s^2+2\tau_2)\big(x_s(\tau_2)-x_s(0)\big)\big(x-x_s(0)\big)+\\
+(\gamma_s^2+2t)\bigg(-\alpha_s^{(M)(2)}(\tau_2)\gamma_s^2 -2\alpha_s^{(M)(2)}(\tau_2)t+2D\gamma_s^2t-2D\gamma_s^2 \tau_2+4D t\tau_2-4D\tau_2^2+\\
+(\gamma_s^2+2t) \big(x_s(\tau_2)-x_s(0)\big)^2\bigg)\Bigg\}\Bigg],
\end{gathered}
\label{resh2}
\end{equation}
where
\begin{equation}
\begin{gathered}
k_{(s)1}(t)=a_{(s)1}(t)-\varkappa h_{(s)1}^{(M)}(t),\\
k_{(s)2}(t)=\dac{1}{2}a_{(s)2}(t)-\dac{\varkappa}{2} h_{(s)2}^{(M)}(t).
\end{gathered}
\label{reshk1}
\end{equation}

The zeroth order EES reads
\begin{equation}
\begin{gathered}
\dot{\sigma}_1^{(0)}=a_{(1)}\sigma_1^{(0)}-\varkappa\sigma_1^{(0)}\left[\sigma_1^{(0)} b_{(1)}+\sigma_2^{(0)} b_{(1,2)}\right],\\
\dot{\sigma}_2^{(0)}=a_{(2)}\sigma_2^{(0)}-\varkappa\sigma_2^{(0)}\left[\sigma_2^{(0)} b_{(2)}+\sigma_1^{(0)} b_{(2,1)}\right].
\end{gathered}
\label{ees0}
\end{equation}
The system \eqref{ees0} is known as the Volterra--Gause equations \cite{gause2003}. This system determines the leading term of asymptotics for $\sigma_s(t)$. Its integrability is necessary for the integrability of the $M$-th order EES. If $b_{(1,2)}=b_{(2,1)}=b_{(1)}=b_{(2)}$ and $a_{(1)}=a_{(2)}$, one can readily get the integral for \eqref{ees0} given by:
\begin{equation}
\dac{d}{dt}\big(\sigma_1^{(0)}+\sigma_2^{(0)}\big)=a_{(1)}\big(\sigma_1^{(0)}+\sigma_2^{(0)}\big)-\varkappa b_{(1)} \big(\sigma_1^{(0)}+\sigma_2^{(0)}\big)^2.
\label{ees1}
\end{equation}
In particular, for $a(x,t)=a(t)$ and $b(x,y,t)=b(t)$, the system \eqref{ees0} is integrable and $u_s(x,t)=v_s^{(0)}(x,t)$, i.e. our method yields the exact solution to \eqref{eq0}, \eqref{dec-1}. In the simplest case $a(t)=\const=a$ and $b(t)=\const=1$ (the multiplier to $b$ can be varied by the nonlinearity factor $\varkappa$), we have
\begin{equation}
\sigma_s(t)=\dac{a e^{at}\sigma_s(0)}{a+\varkappa\big(e^{at}-1\big)\big(\sigma_1(0)+\sigma_2(0)\big)}, \qquad \sigma_s(0)=N_s\gamma_s\sqrt{2\pi D}.
\label{ees2}
\end{equation}
Then, the exact solution $u(x,t)=v_1^{(0)}(x,t)+v_2^{(0)}(x,t)$ reads
\begin{equation}
u(x,t)=\dac{a e^{at}}{a+\varkappa\big(e^{at}-1\big)\big(\sigma_1(0)+\sigma_2(0)\big)}\sum_{s=1,2}\dac{N_s\gamma_s}{\sqrt{2t+\gamma_s^2}}\exp\!\left[-\dac{(x-x_s(0))^2}{4D t + 2 D\gamma_s^2}\right].
\label{ees3}
\end{equation}

\begin{figure}[h]
\centering\begin{minipage}[b][][b]{0.49\linewidth}\centering
    \includegraphics[width=7.5 cm]{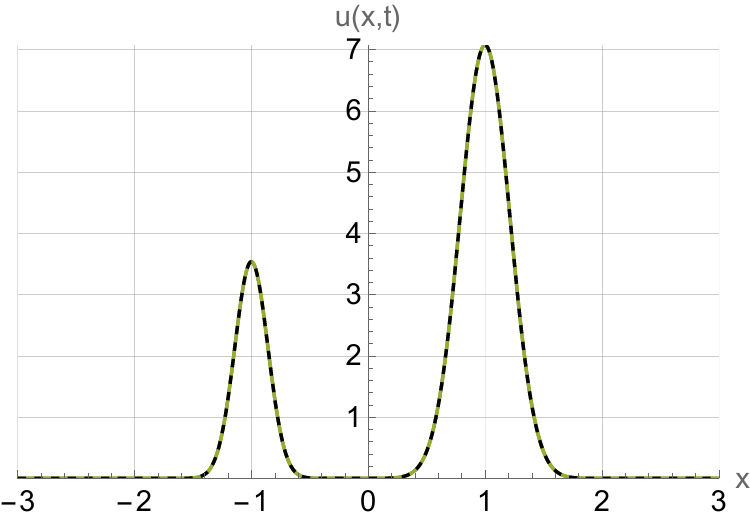} \\ a) $t=0$
  \end{minipage}
 \begin{minipage}[b][][b]{0.49\linewidth} \centering
    \includegraphics[width=7.5 cm]{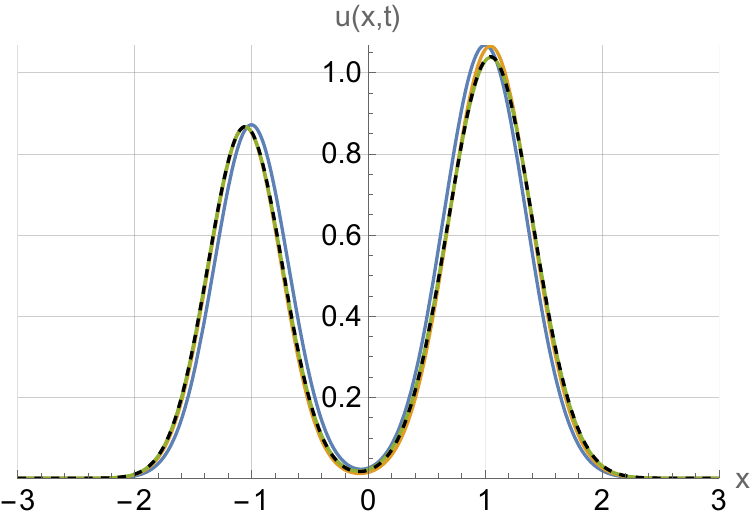} \\ b) $t=2$
  \end{minipage}\\
  \centering\begin{minipage}[b][][b]{0.49\linewidth}\centering
    \includegraphics[width=7.5 cm]{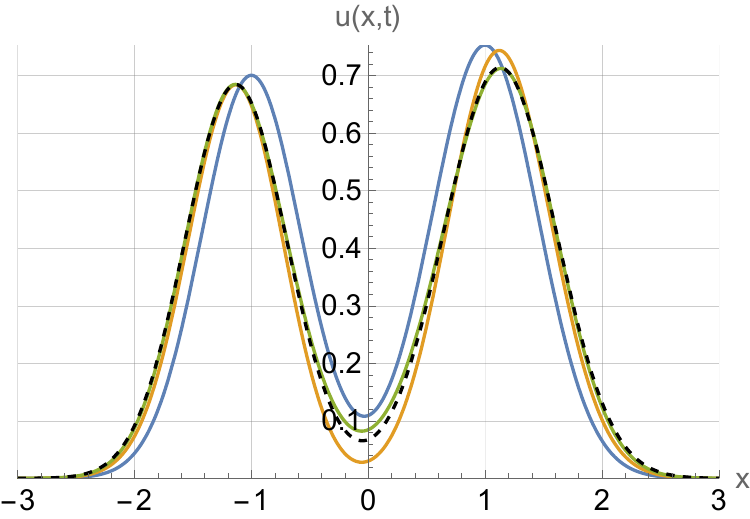} \\ a) $t=4$
  \end{minipage}
 \begin{minipage}[b][][b]{0.49\linewidth} \centering
    \includegraphics[width=7.5 cm]{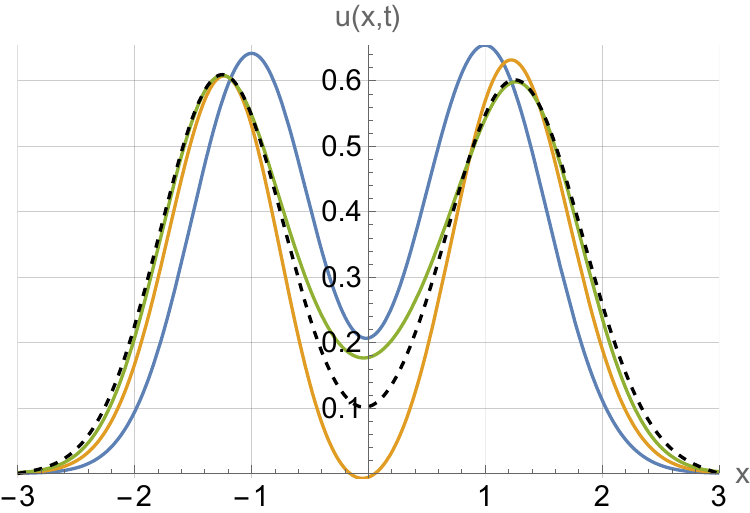} \\ b) $t=6$
  \end{minipage}\\
  \caption{Dependence of $u(x,t)$ on $x$ for various $t$ when the reproduction rate is constant. The blue line is for $u^{(0)}$ (leading term of asymptotics), yellow line is for $u^{(1)}$, green line is for $u^{(2)}$, and black dashed line is for $u_{num}$. \label{fig1}}
\end{figure}

Let us demonstrate our formalism with the less trivial case, when our method yields the asymptotic solution to the non-integrable FKPP. The case $b(x,y,t)=\const$ can be treated as strongly nonlocal FKPP. In order to consider the opposite case, we will consider the kernel $b(x,y,t)$ given by
\begin{equation}
b(x,y,t)=b(x-y)=\exp\Big[-\dac{(x-y)^2}{\zeta^2}\Big].
\label{prim1}
\end{equation}
The initial condition to $u_s(x,t)$ is still given by \eqref{nontr3} and $a(x,t)=\const=a$. It is clear that this example degenerates to $b(x,y,t)=1$ for $\zeta \to \infty$. Firstly, it means that \eqref{ees3} yields the asymptotic solution as $\zeta \to \infty$. Secondly, it means that in this specific example the higher approximations $v_s^{(1)}(x,t)$, $v_s^{(2)}(x,t)$, etc., diminish not only as $D \to 0$ but also as $\zeta\to \infty$. The EES of the second order ($M=2$) is as follows:
\begin{equation}
\begin{gathered}
\dot{X}_1(t)=\dac{\varkappa}{2\zeta^2} \sigma_2(t)\alpha_1^{(2)}(t)\big(X_1(t)-X_2(t)\big) \cdot b\big(X_1(t)-X_2(t)\big),\\
\dot{X}_2(t)=-\dac{\varkappa}{2\zeta^2} \sigma_1(t)\alpha_2^{(2)}(t)\big(X_1(t)-X_2(t)\big) \cdot b\big(X_1(t)-X_2(t)\big),\\
\dot{\sigma}_s(t)=a\sigma_s(t)-\varkappa \sigma_s^2(t) \Big(1-\dac{2\alpha_s^{(2)}(t)}{\zeta^2}\Big)-\\
-\varkappa \sigma_1(t)\sigma_2(t) b\big(X_1(t)-X_2(t)\big) \bigg(1-\dac{\alpha_1^{(2)}(t)+\alpha_2^{(2)}(t)}{\zeta^4}\Big(2\big(X_1(t)-X_2(t)\big)^2-\zeta^2\Big)\bigg),\\
s=1,2.
\end{gathered}
\label{ees4}
\end{equation}
The system \eqref{ees4} of ODEs was solved numerically and its solutions were substituted into \eqref{resh0}, \eqref{resh1}, \eqref{resh2}. Asymptotic solutions to \eqref{eq0} are given by
\begin{equation}
u^{(K)}(x,t)=\sum_{s=1,2}\sum_{k=0}^{K} D^{k/2}v_s^{(k)}(x,t).
\label{orsol1}
\end{equation}
The asymptotic solution correspoding to $K=0$ is the leading term of asymptotics while $K=1$ and $K=2$ yield the asymptotic solutions with higher approximations. Note that the accuracy of the solution \eqref{orsol1} does not increase for $K>M$ where $M=2$ for \eqref{ees4}. Also, we constructed the numerical solution to \eqref{eq0} denoted by $u_{num}(x,t)$ in order to compare it with asymptotic solutions. In Fig. \ref{fig1}, the asymptotic and numerical solutions are given for $D=0.02$, $\varkappa=1$, $\zeta=2$, $a=1$, $N_1=0.5$, $N_2=1$, $x_1(0)=-1$, $x_2(0)=1$, $\gamma_1=1$, $\gamma_2=1.5$.

Fig. \ref{fig1}a shows the initial condition that was exact for both the numerical and asymptotic solution. The original equation \eqref{eq0} has the symmetry $x\to -x$ in our example but the initial condition does not have it. During the evolution of the initial state the function $u(x,t)$ slowly acquires this symmetry. This effect is accurately catched by our asymptotics, especially by $u^{(2)}$ that is very close to the numerical solution $u_{num}$. The difference between the asymptotic solution and the numerical solution grows with a time that is typical for a semiclassical approximation.

\begin{figure}[h]
\centering\begin{minipage}[b][][b]{0.49\linewidth}\centering
    \includegraphics[width=7.5 cm]{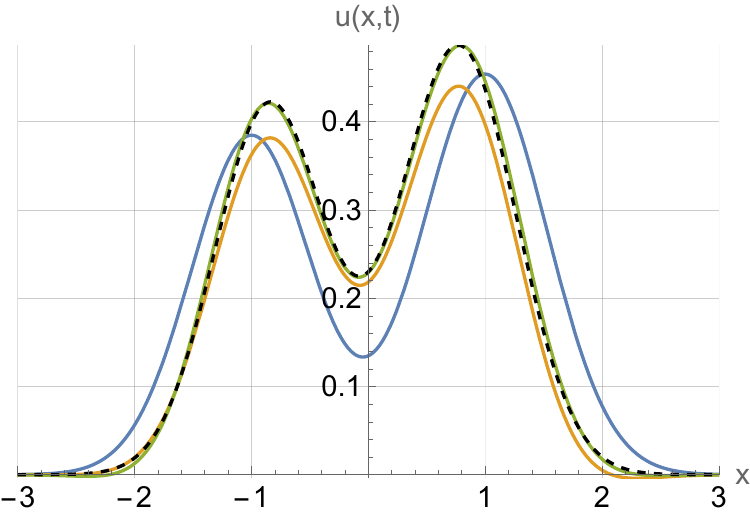} \\ a) $t=6$
  \end{minipage}
 \begin{minipage}[b][][b]{0.49\linewidth} \centering
    \includegraphics[width=7.5 cm]{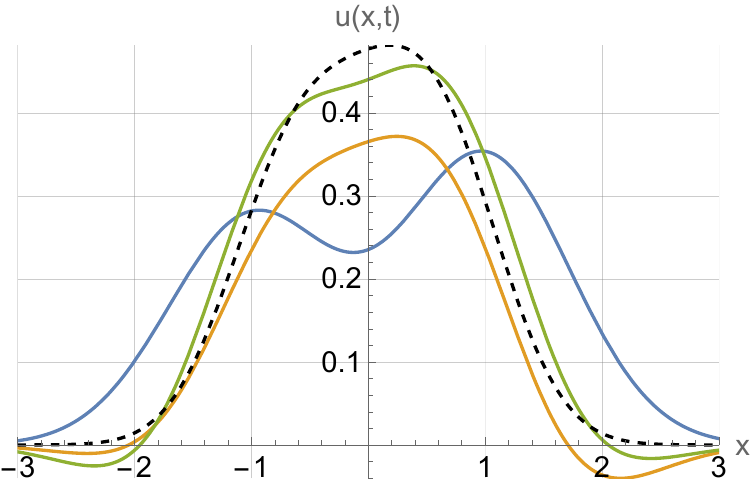} \\ b) $t=12$
  \end{minipage}\\
  \caption{Dependence of $u(x,t)$ on $x$ for various $t$ when the reproduction rate depends on $x$. The blue line is for $u^{(0)}$ (leading term of asymptotics), yellow line is for $u^{(1)}$, green line is for $u^{(2)}$, and black dashed line is for $u_{num}$. \label{fig2}}
\end{figure}

Now, let us study the other case when the population can reproduce only in the limited area, i.e. $a(x,t)>0$ in a closed domain and $a(x,t)<0$ outside this domain. We will consider $a(x,t)=a(x)=1-(x/2)^2$. Other parameters of the equation and initial condition remain the same. In this case, the system \eqref{ees4} reads
\begin{equation}
\begin{gathered}
\dot{X}_1(t)=-\frac{1}{2}X_1(t)\alpha_1^{(2)}(t)+\dac{\varkappa}{2\zeta^2} \sigma_2(t)\alpha_1^{(2)}(t)\big(X_1(t)-X_2(t)\big) \cdot b\big(X_1(t)-X_2(t)\big),\\
\dot{X}_2(t)=-\frac{1}{2}X_2(t)\alpha_2^{(2)}(t)+-\dac{\varkappa}{2\zeta^2} \sigma_1(t)\alpha_2^{(2)}(t)\big(X_1(t)-X_2(t)\big) \cdot b\big(X_1(t)-X_2(t)\big),\\
\dot{\sigma}_s(t)=\Big(a\big(X_s(t)\big)-\frac{1}{4}\alpha_s^{(2)}(t)\Big)\sigma_s(t)-\varkappa \sigma_s^2(t) \Big(1-\dac{2\alpha_s^{(2)}(t)}{\zeta^2}\Big)-\\
-\varkappa \sigma_1(t)\sigma_2(t) b\big(X_1(t)-X_2(t)\big) \bigg(1-\dac{\alpha_1^{(2)}(t)+\alpha_2^{(2)}(t)}{\zeta^4}\Big(2\big(X_1(t)-X_2(t)\big)^2-\zeta^2\Big)\bigg),\\
s=1,2.
\end{gathered}
\label{ees4}
\end{equation}
Fig. \ref{fig2} shows the evolution of the initial state for this case.

As opposite to the previous example, two populations merge into the large one. Asymptotic solutions describe well this merging in spite of the fact that the quasiparticles are hardly distinguishable for $t=12$. Note that Fig. \ref{fig2} shows that the asymptotic solutions obtain the small negative values for some $x$ and large $t$. The original equation \eqref{eq0} and equations \eqref{dec-2} are quasilinear homogeneous equation. Hence, they conserve non-negativity of $u(x,t)$ and $u_s(x,t)$ if $u(x,0)>0$ and $u_s(x,0)>0$. The equation \eqref{als-6a} that determines the leading term of asymptotic is linear homogeneous equation. Hence, $u^{(0)}(x,t)>0$ for any $t$ if $u^{(0)}(x,0)>0$ similar to exact solutions. However, the equations \eqref{als-6b}---\eqref{als-6c} are nonhomogeneous linear equations. It means that $u^{(k)}(x,t)$, $k\geq 1$, can change its sign with a time. It happens when the error of the asymptotic solution becomes large enough. Fig. \ref{fig1} and \ref{fig2} show that the higher approximations greatly increase the accuracy of the asymptotic solutions for large $t$ despite the mentioned defect associated with negative values of higher approximations. It means that the effective time interval where asymptotics are reasonable can probably be substantially prolonged by the increase in the number of accounted terms $v_s^{(k)}(x,t)$.

\section{Conslusion}

\label{sec:con}

We have proposed the method of constructing the asymptotically localized solutions to the Cauchy problem for the nonlocal one-dimensional FKPP equation. The attractive feature of the solutions constructed in this work is that they are localized in a neighbourhood of two trajectories and behave as two quasiparticles, i.e. this method allows one not only describe the simple localized population but the evolution of a pattern. The asymptotics have been constructed within the weak diffusion approximation. The Cauchy problem for the original nonlocal FKPP equation is reduced to the Cauchy problem for the dynamical system of ODEs. The proposed method allows one to construct the solutions with any accuracy with respect to the small diffusion parameter $D$ on a limited time interval. The asymptotic solutions to the FKPP equation have been found among the solutions to the associated linear equations \eqref{als-6a}--\eqref{als-6d}.

We have derived the explicit expression for the first three terms of the asymptotic expansion and studied them for various cases. In the first case with the constant reproduction rate and strongly nonlocal nonlinearity, our method yields the exact solutions to the nonlocal FKPP. Then, we have considered two more complex cases with non-symmetrical initial condition when two quasiparticle correspond to the population of different sizes. In these cases, the evolution of the initial condition leads to the equalization of these population to the symmetric one, i.e. the solutions obtain the asymptotic symmetry. After this fast process, the solution behave differently depending on the chosen equation coefficients. When the reproduction rate is constant, the quasiparticles diverge in different directions. When the reproduction rate is localized in a certain area, the quasiparticles merge into large one. For these two cases, we have compared the asymptotic solutions with the numerical ones. Such comparison shows that our asymptotic solutions accurately describe the effects under consideration. Therefore, our approach can be helpful for the prediction of population dynamics within the framework of the nonlocal FKPP model. The proposed ideas can be naturally generalized to the $n$-dimensional FKPP equation and greater number of quasiparticles.

Note that the case of a single quasiparticle in the FKPP equation can also be treated as a particular case of work \cite{kulagin2023semiclassical} in a manner. We are encouraged to generalized the method \cite{kulagin2023semiclassical} for few quasiparticles using ideas of this work.

\clearpage

\section*{Acknowledgement}

The study is supported by Russian Science Foundation, project no. 23-71-01047, https://rscf.ru/en/project/23-71-01047/.

\bibliography{lit1}

\end{document}